\documentclass[aps,twocolumn]{revtex4}
\usepackage{graphics}
\begin{document}
\newcommand{\bstfile}{aps} %alternative styles: osa, prasty or revtex
\newcommand{\bibs}{c:/Bibliography/final}
\draft
\title{Pushing the hyperpolarizability to the limit}
\author{Juefei Zhou, Mark G. Kuzyk, and David S. Watkins*}
\address{Department of Physics and Astronomy; and Mathematics*, Washington State University, Pullman, Washington  99164-2814}
\date{\today}

\begin{abstract}
We use numerical optimization to find a one-dimensional potential
energy function that yields the largest hyperpolarizability, which
we find is within 30\% of the fundamental limit.  Our results reveal
insights into the character of the potential energy functions and
wavefunctions that lead to the largest hyperpolarizability.  We
suggest that donor-acceptor molecules with a conjugated bridge with
many sites of reduced conjugation to impart conjugation modulation
may be the best paradigm for making materials with huge
hyperpolarizabilities that approach the fundamental limit.
\end{abstract}

\maketitle

The study of nonlinear optical materials has been driven by the
quest for making the largest possible nonlinear response in a medium
for use in a broad range of optical applications such as
telecommunications,\cite{wang04.01} optical data
storage,\cite{Olson02.01} three-dimensional
nano-photolithography,\cite{cumps99.01,kawat01.01} and making new
materials\cite{karot04.01} for novel cancer
therapies.\cite{roy03.01}

A natural questions arises whether there is a limit to the nonlinear
response and if so, how does one make a molecule to get to the
limit.  The former has been answered in the affirmative using sum
rules,\cite{kuzyk00.02,kuzyk01.01,kuzyk00.01,kuzyk03.01,kuzyk03.02,kuzyk04.02}
and the fundamental limit is given by,
\begin{equation}\label{limit}
\beta_{MAX} = \sqrt[4]{3} \left( \frac {e \hbar} {\sqrt{m}}
\right)^3 \cdot \frac {N^{3/2}} {E_{10}^{7/2}} ,
\end{equation}
where $N$ is the number of electrons and $E_{10}$ the energy to the
first excited state.  This result has not yielded clear guidance on
how to design better molecules, but rather has set the bar for
assessing the performance of molecules. In fact, the largest
nonlinear susceptibilities of the best molecules fall short of the
fundamental limit by a factor of
$10^{3/2}$.\cite{kuzyk03.02,Kuzyk03.05,Tripa04.01}  A
Sum-Over-States (SOS) calculation of the
hyperpolarizability\cite{orr71.01} using the analytical
wavefunctions of the clipped harmonic oscillator yields a value that
is within a factor of 2 of the limit.\cite{Tripa04.01} Thus, the
factor-of-thirty gap between the fundamental limit and the apparent
upper bound of the best molecules is not of a fundamental nature.
So, it should be possible in principle to make materials with
second-order susceptibilities that are a factor of 30 bigger than
the currently best materials.  In this letter, we report on the
character of the potential energy function of a one-dimensional
system that yields a hyperpolarizability, $\beta$, that is within
30\% of the fundamental limit, and use this potential to propose a
new paradigm for molecular engineering.

There are two equivalent expressions for $\beta$.  The standard one,
$\beta_{SOS}$, as calculated by Orr and Ward,\cite{orr71.01} and the
dipole-free expression, $\beta_{DF}$.\cite{kuzyk05.02}  The standard
one is overspecified in the sense that it is possible to pick
unphysical values of the energies and matrix elements, which violate
the precepts of quantum mechanics.  The dipole free expression, in
contrast, is simplified into a reduced form that contains no dipole
terms.  Since the two expressions should yield the same results if
all the matrix elements are consistent with the sum rules,
deviations between them can be used as a convergence
test.\cite{kuzyk05.02} However, we later discuss how to treat
exceptions where $\beta_{DF}$ is accurate and $\beta_{SOS}$ is not.

Our approach is to numerically calculate $\beta$ for a specific
potential, then, to use an optimization algorithm that continuously
varies the potential in a way that maximizes $\beta$ - using the
convergence test to determine if the result is reliable. Our code is
written in MATLAB.  For each trial potential we use a quadratic
finite element method \cite{zienk05.01} to approximate the
Schr\"{o}dinger eigenvalue problem and the implicitly restarted
Arnoldi method \cite{soren92.01} to compute the wave functions and
energy levels.  To optimize $\beta$ we use the Nelder-Mead simplex
algorithm \cite{lagar98.01}.

\begin{figure}
\includegraphics{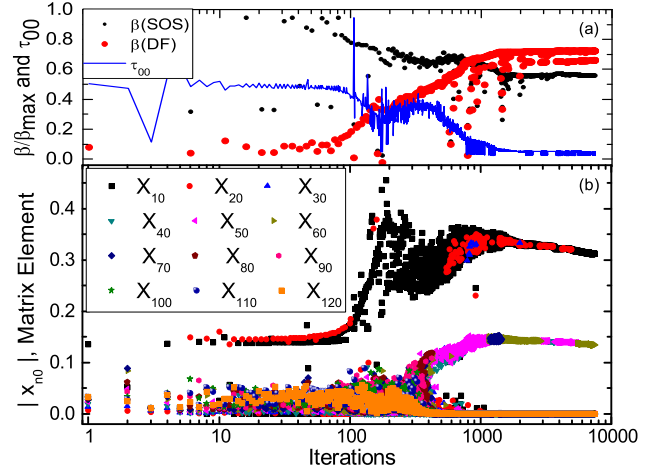}
\caption{Evolution of (a) $\beta$ and $\tau_{00}$; and (b) $x_{n0}$
as a function of number of iterations of optimization.}
\label{fig:IterationSummary}
\end{figure}
Figure \ref{fig:IterationSummary}a shows the evolution of the
hyperpolarizability as a function of the number of iterations of the
optimization algorithm, applied to $\beta_{DF}$.  All
hyperpolarizabilities are normalized to the fundamental limit.  The
initial potential energy function is a hyperbolic tangent with
infinite potential at the boundaries.  We select this potential
because it is localized near the origin, yet flat elsewhere,
allowing the optimization process to work without any initial
biases.  Also, it meets the criteria of being noncentrosymmetric, as
required for nonvanishing hyperpolarizability.  After 7,000
iterations, the algorithm converges to just over
$\beta_{DF}/\beta_{MAX} = 0.72$, the largest hyperpolarizability
seen to date.  No other starting potential, including polynomials,
and fractional exponents, leads to larger $\beta_{DF}$.

As a second convergence test, we use the fractional deviation from
the ground state sum rule, which in terms of $\kappa_{00}$ as
defined in the literature, is given by,\cite{kuzyk06.01}
\begin{equation}
\tau_{00} = 1 - \sum_n \frac {E_{n0}} {E_{10}} \left| \frac {x_{n0}}
{x_{10}^{max}} \right|^2 \equiv 1 - \kappa_{00}, \label{tau00}
\end{equation}
where $E_{n0}$ is the energy difference between state $n$ and $0$,
$x_{nm}$ is the position matrix element between state $n$ and $m$,
and $x_{10}^{max}$ is the fundamental limit of $x_{10}$. $\tau_{00}$
approaches zero when the optimization process yields maximum
$\beta_{DF}$, supporting the fact that the numerical computations
accurately represent the system.  When, on the other hand,
$\beta_{SOS}$ is optimized, $\tau_{00}$ is large and the two forms
of $\beta$ diverge appreciably, illustrating the robustness of the
dipole-free form and the pitfalls of using $\beta_{SOS}$ for
optimization.  Our calculations of $\beta$ include a total of 15
states, so a large value of $\tau_{00}$ is more of an indication
that not enough states are being used in $\beta_{SOS}$ rather than
inaccuracies of the wave functions.  The fact that $\beta_{SOS}$ and
$\beta_{DF}$ do not converge illustrates the need for two
independent convergence tests.

Figure \ref{fig:IterationSummary}b shows the evolution of the
position matrix elements.  When $\beta_{DF}$ is small, many of these
matrix elements are non-negligible, so many states will contribute
to $\beta$.  By a thousand iterations when the hyperpolarizability
approaches its largest value, all matrix elements vanish except for
two of them.  This clearly shows that as the fundamental limit of
$\beta$ is approached, the system collapses to a three-level model
for $\beta$.  This is consistent with the three-level ansatz,
previously proposed, that a three-level model describes a system
with a hyperpolarizability at the fundamental
limit.\cite{kuzyk05.01}  There is mounting evidence that the
three-level ansatz also is obeyed in planar
molecules.\cite{kuzyk06.02a}  We stress that our results do not
prove the converse; that all molecules with hyperpolarizabilities at
the fundamental limit must be represented by a three-level model.
However, since (1) the three-level ansatz is used to determine the
fundamental limits, and, all reliable measurements and calculations
yield hyperpolarizabilities below the fundamental limit; and (2)
numerical optimization yields a three-level model, this form of
induction provides strong support for the ansatz.

\begin{figure}
\includegraphics{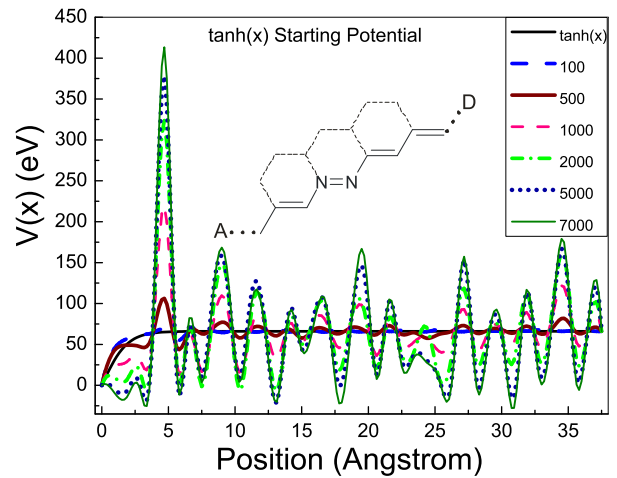}
\caption{Evolution of potential energy as a function iteration.
Inset shows a class of proposed molecules, where ``..." refers to a continuation of the bridge using the same theme.}
\label{fig:PotentialEvolution}
\end{figure}
Figure \ref{fig:PotentialEvolution} shows the evolution of the
potential energy function for a length scale on the order of a large
molecule. Optimization of the hyperpolarizability clearly favors a
potential with large oscillations. Interestingly, all starting
potentials we have studied, independent of their initial form - such
as polynomials, power laws, and piecewise continuous functions,
develop such wild oscillations of about the same period when the
hyperpolarizability is maximized. We find that these oscillations
serve to localize each eigenfunction at different positions.  Early
in the optimization process, the wavefunctions are all delocalized
while at 7,000 iterations, most of the wavefunctions are mutually
non-overlapping.

\begin{figure}
\includegraphics{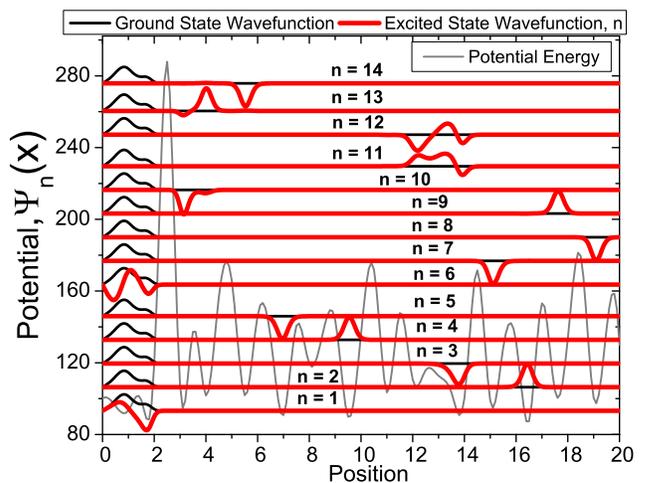}
\caption{Potential energy and energy eigenfunctions after 7,000
iterations, when $\beta_{DF}$ is optimized.}
\label{fig:FinalEigenstates}
\end{figure}
Figure \ref{fig:FinalEigenstates} shows the eigenfunctions and
potential energy function after 7,000 iterations. Only excited
states number 1 and 6 have appreciable overlap with the ground state
and with each other. As such, the only term that contributes to the
hyperpolarizability is proportional to $x_{01} x_{16} x_{60} /
E_{10} E_{60}$.  Note that with other starting potentials, states
other then $0$, $1$, $6$ may be important.  Most significantly,
$\beta$ is maximum for a three-level system.

Under the three-level ansatz, the normalized hyperpolarizability can
be expressed as a product of two functions, $\beta / \beta_{MAX} f(E) G(X)$,\cite{Tripa04.01} where $X = x_{10} / x_{10}^{MAX}$ and
$E = E_{10} / E_{20}$.  ``$1$" and ``$2$" label the states with the
largest transition moments to the ground state.  The function $f(E)$
is maximum when $E=0$ ($f(0)=1$) and $G(X)$ is maximum when
$X=\sqrt[-4]{3}$ ($G(\sqrt[-4]{3}) = 1$).  So, a hyperpolarizability
that approaches the fundamental limit should have two dominant
states that have well separated eigenenergies; and, $x_{10} \sqrt[-4]{3} x_{10}^{MAX}$.  The optimized wavefunctions yield $E 0.314$ and $X=0.775$, so $f(E)=0.892$ and $G(X) = 0.999$.  The
transition moment is near optimal but the energy level spacing can
be improved.  If the system were truly 3 levels, $\beta /
\beta_{MAX} = f(E) G(X) = 0.89$ compared with the optimized value of
$\beta_{DF} / \beta_{MAX} = 0.72$.

One might argue that our results are artificial because we are using
only 15 states, and, because our space is bounded by an infinite
potential.  These two issues go hand in hand. In our calculations,
the potential energy function is represented by a cubic spline of 20
points.  As such, it would be numerically impossible to localize
more than about 15 eigenfunctions. Furthermore, higher-energy
wavefunctions would interact with the walls; and, for high-enough
energies, the wavefunctions would have the character of a particle
in a box.  So, to increase the number of energy levels, one would
need to increase the number of points in the spline, which would
lead to more oscillations that would localize more of the excited
states.  Based on the pattern that we have observed, when increasing
the number of states and making the potential correspondingly
broader, we expect that more non-overlapping wavefunctions would
develop, yielding a similar conclusion.  So, in the limit of
infinitely wide space, the conclusion should be the same.

In summary, we have applied an optimization process to determine
what potential energy function yields the largest value of the
hyperpolarizability.  As a function of iteration, the
hyperpolarizability approaches the fundamental limit, but never
exceeds it.  When $\beta$ converges to the maximal value, the
expression for $\beta$ collapses to a three-level model, supporting
the three-level ansatz.  Strong oscillations in the potential energy
function serve to isolate the wavefunctions to prevent overlap
between all states but three of them.  This behavior suggests that
long linear organic molecules with regions of conjugation
(representing the dips in the potential) interspersed by regions of
reduced conjugation (representing the peaks) and flanked by a donor
and acceptor to break the symmetry, would be a promising new
paradigm.

The inset in Figure \ref{fig:PotentialEvolution} shows one example
of a new molecular paradigm that should exhibit the required
modulation of conjugation over a single and double bond pair spatial
period. The dashed bonds represent the absence or presence of rings
to control the degree of conjugation and/or molecular planarity.
Note that at 100 iterations, the oscillations are on the order of
1eV (common energy scale for molecules), where $\beta/\beta_{MAX} 0.2$ -- well above the apparent limit of $\beta/\beta_{MAX} = 0.03$.
So, our new paradigm should lead to a tenfold improvement over
today's best molecules.  Alternatively, our method could guide the
fabrication of multiple quantum well structures using a variety of
layered organic molecules that impart to it an oscillating potential
energy profile.

We  plan to apply our general method to higher-order
nonlinear-optical susceptibilities.  In addition, it can be applied
to resonant processes and to any general combinations of laser
wavelengths.  With regards to the second hyperpolarizability,
$\gamma$, since past calculations suggest that the three-level
ansatz also should hold, we would expect the same sort of
conclusion.  Should this approach be successful, new and exceptional
materials could be developed to make more efficient devices and
novel applications.

{\bf Acknowledgements: } MGK thanks the National Science Foundation
(ECS-0354736) and Wright Paterson Air Force Base for generously
supporting this work.

%\bibliography{\bibs}

\begin{thebibliography}{22}
\expandafter\ifx\csname
natexlab\endcsname\relax\def\natexlab#1{#1}\fi
\expandafter\ifx\csname bibnamefont\endcsname\relax
  \def\bibnamefont#1{#1}\fi
\expandafter\ifx\csname bibfnamefont\endcsname\relax
  \def\bibfnamefont#1{#1}\fi
\expandafter\ifx\csname citenamefont\endcsname\relax
  \def\citenamefont#1{#1}\fi
\expandafter\ifx\csname url\endcsname\relax
  \def\url#1{\texttt{#1}}\fi
\expandafter\ifx\csname urlprefix\endcsname\relax\def\urlprefix{URL
}\fi \providecommand{\bibinfo}[2]{#2}
\providecommand{\eprint}[2][]{\url{#2}}

\bibitem[{\citenamefont{Chen et~al.}(2004)\citenamefont{Chen, Kuang, Wang, and
  Sargent}}]{wang04.01}
\bibinfo{author}{\bibfnamefont{Q.~Y.} \bibnamefont{Chen}},
  \bibinfo{author}{\bibfnamefont{L.}~\bibnamefont{Kuang}},
  \bibinfo{author}{\bibfnamefont{Z.~Y.} \bibnamefont{Wang}}, \bibnamefont{and}
  \bibinfo{author}{\bibfnamefont{E.~H.} \bibnamefont{Sargent}},
  \bibinfo{journal}{Nano. Lett.} \textbf{\bibinfo{volume}{4}},
  \bibinfo{pages}{1673} (\bibinfo{year}{2004}).

\bibitem[{\citenamefont{Olson et~al.}(2002)\citenamefont{Olson, Previte, and
  Fourkas}}]{Olson02.01}
\bibinfo{author}{\bibfnamefont{C.~E.} \bibnamefont{Olson}},
  \bibinfo{author}{\bibfnamefont{M.~J.~R.} \bibnamefont{Previte}},
  \bibnamefont{and} \bibinfo{author}{\bibfnamefont{J.~T.}
  \bibnamefont{Fourkas}}, \bibinfo{journal}{Nature Materials}
  \textbf{\bibinfo{volume}{1}}, \bibinfo{pages}{225} (\bibinfo{year}{2002}).

\bibitem[{\citenamefont{Cumpston et~al.}(1999)\citenamefont{Cumpston,
  Ananthavel, Barlow, Dyer, Ehrlich, Erskine, Heikal, Kuebler, Lee,
  McCord-Maughon, Qin, R\"{o}ckel, Rumi, Wu, Marder and Perry}}]{cumps99.01}
\bibinfo{author}{\bibfnamefont{B.~H.} \bibnamefont{Cumpston}},
  \bibinfo{author}{\bibfnamefont{S.~P.} \bibnamefont{Ananthavel}},
  \bibinfo{author}{\bibfnamefont{S.}~\bibnamefont{Barlow}},
  \bibinfo{author}{\bibfnamefont{D.~L.} \bibnamefont{Dyer}},
  \bibinfo{author}{\bibfnamefont{J.~E.} \bibnamefont{Ehrlich}},
  \bibinfo{author}{\bibfnamefont{L.~L.} \bibnamefont{Erskine}},
  \bibinfo{author}{\bibfnamefont{A.~A.} \bibnamefont{Heikal}},
  \bibinfo{author}{\bibfnamefont{S.~M.} \bibnamefont{Kuebler}},
  \bibinfo{author}{\bibfnamefont{I.-Y.~S.} \bibnamefont{Lee}},
   \bibinfo{author}{\bibfnamefont{D.}~\bibnamefont{McCord-Maughon}},
   \bibinfo{author}{\bibfnamefont{J.}~\bibnamefont{Qin}},
   \bibinfo{author}{\bibfnamefont{H.}~\bibnamefont{R\"{o}ckel}},
   \bibinfo{author}{\bibfnamefont{M.}~\bibnamefont{Rumi}},
   \bibinfo{author}{\bibfnamefont{X.-L.}~\bibnamefont{Wu}},
   \bibinfo{author}{\bibfnamefont{S.~R.}~\bibnamefont{Marder}},
   \bibnamefont{and} \bibinfo{author}{\bibfnamefont{J.~W.}~\bibnamefont{Perry}},
  \bibinfo{journal}{Nature}
  \textbf{\bibinfo{volume}{398}}, \bibinfo{pages}{51} (\bibinfo{year}{1999}).

\bibitem[{\citenamefont{Kawata et~al.}(2001)\citenamefont{Kawata, Sun, Tanaka,
  and Takada}}]{kawat01.01}
\bibinfo{author}{\bibfnamefont{S.}~\bibnamefont{Kawata}},
  \bibinfo{author}{\bibfnamefont{H.-B.} \bibnamefont{Sun}},
  \bibinfo{author}{\bibfnamefont{T.}~\bibnamefont{Tanaka}}, \bibnamefont{and}
  \bibinfo{author}{\bibfnamefont{K.}~\bibnamefont{Takada}},
  \bibinfo{journal}{Nature} \textbf{\bibinfo{volume}{412}},
  \bibinfo{pages}{697} (\bibinfo{year}{2001}).

\bibitem[{\citenamefont{Karotki et~al.}(2004)\citenamefont{Karotki, Drobizhev,
  Dzenis, Taylor, Anderson, and Rebane}}]{karot04.01}
\bibinfo{author}{\bibfnamefont{A.}~\bibnamefont{Karotki}},
  \bibinfo{author}{\bibfnamefont{M.}~\bibnamefont{Drobizhev}},
  \bibinfo{author}{\bibfnamefont{Y.}~\bibnamefont{Dzenis}},
  \bibinfo{author}{\bibfnamefont{P.~N.} \bibnamefont{Taylor}},
  \bibinfo{author}{\bibfnamefont{H.~L.} \bibnamefont{Anderson}},
  \bibnamefont{and} \bibinfo{author}{\bibfnamefont{A.}~\bibnamefont{Rebane}},
  \bibinfo{journal}{Phys. Chem. Chem. Phys.} \textbf{\bibinfo{volume}{6}},
  \bibinfo{pages}{7} (\bibinfo{year}{2004}).

\bibitem[{\citenamefont{Roy et~al.}(2003)\citenamefont{Roy, Y., Pudavar,
  Bergey, Oseroff, Morgan, Dougherty, and Prasad}}]{roy03.01}
\bibinfo{author}{\bibfnamefont{I.}~\bibnamefont{Roy}},
  \bibinfo{author}{\bibfnamefont{O.~T.} \bibnamefont{Y.}},
  \bibinfo{author}{\bibfnamefont{H.~E.} \bibnamefont{Pudavar}},
  \bibinfo{author}{\bibfnamefont{E.~J.} \bibnamefont{Bergey}},
  \bibinfo{author}{\bibfnamefont{A.~R.} \bibnamefont{Oseroff}},
  \bibinfo{author}{\bibfnamefont{J.}~\bibnamefont{Morgan}},
  \bibinfo{author}{\bibfnamefont{T.~J.} \bibnamefont{Dougherty}},
  \bibnamefont{and} \bibinfo{author}{\bibfnamefont{P.~N.}
  \bibnamefont{Prasad}}, \bibinfo{journal}{J. Am. Chem. Soc.}
  \textbf{\bibinfo{volume}{125}}, \bibinfo{pages}{7860} (\bibinfo{year}{2003}).

\bibitem[{\citenamefont{Kuzyk}(2000{\natexlab{a}})}]{kuzyk00.02}
\bibinfo{author}{\bibfnamefont{M.~G.} \bibnamefont{Kuzyk}},
  \bibinfo{journal}{Opt. Lett.} \textbf{\bibinfo{volume}{25}},
  \bibinfo{pages}{1183} (\bibinfo{year}{2000}{\natexlab{a}}).

\bibitem[{\citenamefont{Kuzyk}(2001)}]{kuzyk01.01}
\bibinfo{author}{\bibfnamefont{M.~G.} \bibnamefont{Kuzyk}},
  \bibinfo{journal}{IEEE Journal on Selected Topics in Quantum Electronics}
  \textbf{\bibinfo{volume}{7}}, \bibinfo{pages}{774 } (\bibinfo{year}{2001}).

\bibitem[{\citenamefont{Kuzyk}(2000{\natexlab{b}})}]{kuzyk00.01}
\bibinfo{author}{\bibfnamefont{M.~G.} \bibnamefont{Kuzyk}},
  \bibinfo{journal}{Phys. Rev. Lett.} \textbf{\bibinfo{volume}{85}},
  \bibinfo{pages}{1218} (\bibinfo{year}{2000}{\natexlab{b}}).

\bibitem[{\citenamefont{Kuzyk}(2003{\natexlab{a}})}]{kuzyk03.01}
\bibinfo{author}{\bibfnamefont{M.~G.} \bibnamefont{Kuzyk}},
  \bibinfo{journal}{Opt. Lett.} \textbf{\bibinfo{volume}{28}},
  \bibinfo{pages}{135} (\bibinfo{year}{2003}{\natexlab{a}}).

\bibitem[{\citenamefont{Kuzyk}(2003{\natexlab{b}})}]{kuzyk03.02}
\bibinfo{author}{\bibfnamefont{M.~G.} \bibnamefont{Kuzyk}},
  \bibinfo{journal}{Phys. Rev. Lett.} \textbf{\bibinfo{volume}{90}},
  \bibinfo{pages}{039902} (\bibinfo{year}{2003}{\natexlab{b}}).

\bibitem[{\citenamefont{Kuzyk}(2004)}]{kuzyk04.02}
\bibinfo{author}{\bibfnamefont{M.~G.} \bibnamefont{Kuzyk}},
  \bibinfo{journal}{J. Nonl. Opt. Phys. \& Mat.} \textbf{\bibinfo{volume}{13}},
  \bibinfo{pages}{461} (\bibinfo{year}{2004}).

\bibitem[{\citenamefont{Kuzyk}(2003{\natexlab{c}})}]{Kuzyk03.05}
\bibinfo{author}{\bibfnamefont{M.~G.} \bibnamefont{Kuzyk}},
  \bibinfo{journal}{Optics \& Photonics News} \textbf{\bibinfo{volume}{14}},
  \bibinfo{pages}{26} (\bibinfo{year}{2003}{\natexlab{c}}).

\bibitem[{\citenamefont{Tripathi et~al.}(2004)\citenamefont{Tripathi,
  P\'{e}rez~Moreno, Kuzyk, Coe, Clays, and Kelley}}]{Tripa04.01}
\bibinfo{author}{\bibfnamefont{K.}~\bibnamefont{Tripathi}},
  \bibinfo{author}{\bibfnamefont{J.}~\bibnamefont{P\'{e}rez~Moreno}},
  \bibinfo{author}{\bibfnamefont{M.~G.} \bibnamefont{Kuzyk}},
  \bibinfo{author}{\bibfnamefont{B.~J.} \bibnamefont{Coe}},
  \bibinfo{author}{\bibfnamefont{K.}~\bibnamefont{Clays}}, \bibnamefont{and}
  \bibinfo{author}{\bibfnamefont{A.~M.} \bibnamefont{Kelley}},
  \bibinfo{journal}{J. Chem. Phys.} \textbf{\bibinfo{volume}{121}},
  \bibinfo{pages}{7932} (\bibinfo{year}{2004}).

\bibitem[{\citenamefont{Orr and Ward}(1971)}]{orr71.01}
\bibinfo{author}{\bibfnamefont{B.~J.} \bibnamefont{Orr}} \bibnamefont{and}
  \bibinfo{author}{\bibfnamefont{J.~F.} \bibnamefont{Ward}},
  \bibinfo{journal}{Molecular Physics} \textbf{\bibinfo{volume}{20}},
  \bibinfo{pages}{513} (\bibinfo{year}{1971}).

\bibitem[{\citenamefont{Kuzyk}(2005{\natexlab{a}})}]{kuzyk05.02}
\bibinfo{author}{\bibfnamefont{M.~G.} \bibnamefont{Kuzyk}},
  \bibinfo{journal}{Phys. Rev. A} \textbf{\bibinfo{volume}{72}},
  \bibinfo{pages}{053819} (\bibinfo{year}{2005}{\natexlab{a}}).

\bibitem[{\citenamefont{Zienkiewicz et~al.}(2005)\citenamefont{Zienkiewicz,
  Taylor, and Zhu}}]{zienk05.01}
\bibinfo{author}{\bibfnamefont{O.~C.} \bibnamefont{Zienkiewicz}},
  \bibinfo{author}{\bibfnamefont{R.~L.} \bibnamefont{Taylor}},
  \bibnamefont{and} \bibinfo{author}{\bibfnamefont{J.~Z.} \bibnamefont{Zhu}},
  \emph{\bibinfo{title}{The Finite Element Method: Its Basis and Fundamentals}}
  (\bibinfo{publisher}{Butterworth-Heinemanm}, \bibinfo{year}{2005}),
  \bibinfo{edition}{6th} ed.

\bibitem[{\citenamefont{Sorensen}(1992)}]{soren92.01}
\bibinfo{author}{\bibfnamefont{D.~C.} \bibnamefont{Sorensen}},
  \bibinfo{journal}{{SIAM} J. Matrix Anal. Appl.}
  \textbf{\bibinfo{volume}{13}}, \bibinfo{pages}{357} (\bibinfo{year}{1992}).

\bibitem[{\citenamefont{Lagarias et~al.}(1998)\citenamefont{Lagarias, Reeds,
  Wright, and Wright}}]{lagar98.01}
\bibinfo{author}{\bibfnamefont{J.~C.} \bibnamefont{Lagarias}},
  \bibinfo{author}{\bibfnamefont{J.~A.} \bibnamefont{Reeds}},
  \bibinfo{author}{\bibfnamefont{M.~H.} \bibnamefont{Wright}},
  \bibnamefont{and} \bibinfo{author}{\bibfnamefont{P.}~\bibnamefont{Wright}},
  \bibinfo{journal}{SIAM J. Optim.} \textbf{\bibinfo{volume}{9}},
  \bibinfo{pages}{112} (\bibinfo{year}{1998}).

\bibitem[{\citenamefont{Kuzyk}(2006)}]{kuzyk06.01}
\bibinfo{author}{\bibfnamefont{M.~G.} \bibnamefont{Kuzyk}},
  \bibinfo{journal}{J. Nonl. Opt. Phys. \& Mat.} \textbf{\bibinfo{volume}{15}},
  \bibinfo{pages}{77} (\bibinfo{year}{2006}).

\bibitem[{\citenamefont{Kuzyk}(2005{\natexlab{b}})}]{kuzyk05.01}
\bibinfo{author}{\bibfnamefont{M.~G.} \bibnamefont{Kuzyk}},
  \bibinfo{journal}{Phys. Rev. Lett.} \textbf{\bibinfo{volume}{95}},
  \bibinfo{pages}{109402} (\bibinfo{year}{2005}{\natexlab{b}}).

\bibitem[{\citenamefont{Kuzyk and Watkins}(2006)}]{kuzyk06.02a}
\bibinfo{author}{\bibfnamefont{M.~G.} \bibnamefont{Kuzyk}} \bibnamefont{and}
  \bibinfo{author}{\bibfnamefont{D.~S.} \bibnamefont{Watkins}},
  \bibinfo{journal}{arXiv:physics/0601172}  (\bibinfo{year}{2006}).

\end{thebibliography}

\clearpage

\end{document}